\documentclass[preprint,aps,floatfix]{revtex4-1}
\usepackage{hyperref}
\everymath={\displaystyle}
\usepackage{young}
\usepackage{rotating}
\usepackage{longtable}
\def\1{\mbox{l\hspace{-0.53em}1}}
\newcommand{\fr}{\frac}
\usepackage{multirow}

\newlength{\AccoHaut}

\begin{document}
\title{SU(3) flavor symmetry breaking in large $N_c$ excited hyperons}
\author{Fl. Stancu\footnote{E-mail address: fstancu@ulg.ac.be}}
\affiliation{University of Li\`ege, Institute of Physics B5, Sart Tilman,
B-4000 Li\`ege 1, Belgium}

\date{\today}

\begin{abstract}

The $1/N_c$ expansion method for studying the mass spectrum of excited baryons is shortly reviewed  
together with applications to mixed symmetric states. 
The $[{\bf 70, \ell^+}]$ multiplet, belonging to the $N$ = 2 band,
is reanalyzed, with emphasis on hyperons 
and the SU(3) symmetry breaking operators entering the mass formula to first order.
An important result is that the hierarchy of masses as a function of strangeness is
correctly reproduced for all multiplets.
Predictions for unknown  excited hyperons to SU(6) $\times$ O(3) multiplets are made.

\end{abstract}

\maketitle

\section{Introduction}
Understanding the baryon resonances and their group theory classification is an
essential and current topic in hadronic physics.
It is well known that the number of observed resonances is smaller than 
the number of excited baryons predicted by the quark model. The number of "missing" resonances
is much larger in the strange sector. The question is whether or not
the missing hyperons with strangeness $\mathcal{S}$ = - 1, - 2, - 3 are due to lack of experimental data 
or due to models based on SU(3) symmetry breaking. 
Experimentally, the hyperons are difficult to produce. In particular for $\mathcal{S}$ = - 2 hyperons, kaon-nucleon
or $\Sigma$-hyperon induced reactions are required and the planned kaon beams 
at Thomas Jefferson National Acceleration Facility (JLAB) and the Japan Proton Accelerator Research Complex (J-PARK)
are expected to improve the situation \cite{JLAB}.

Here we discuss a theoretical approach attempting 
to make an SU(3) classification of excited baryons in the framework of the $1/N_c$ expansion method,
where $N_c$ is the number of colors.  
This method, proposed by 't Hooft \cite{'tHooft:1973jz} and applied to baryons by 
Witten  \cite{Witten:1979kh},
is a powerful tool to study baryon spectroscopy. The underlying symmetry 
is SU($2N_f$) which 
results from the discovery
that, for $N_f$ flavors, the ground state baryons
display an exact contracted SU($2N_f$) spin-flavor symmetry in the
large $N_c$ limit of QCD \cite{Gervais:1983wq,DM93}. 
The Skyrme model, the strong coupling theory \cite{Cook:1965qu} and the static quark model share a common
symmetry with QCD baryons in the large $N_c$ limit \cite{Bardakci:1983ev}.
 
The $1/N_c$ expansion method has been applied with great success
to the ground state baryons, described
by the symmetric representation $\bf 56$ of SU(6) 
\cite{DM93,Jenk1,DJM94,DJM95,CGO94,JL95}. 
At $N_c \rightarrow \infty$ the ground state baryons are degenerate.
At large, but finite $N_c$, the mass splitting 
starts at order $1/N_c$ as first observed in Ref. \cite{Bardakci:1983ev}.
For a review regarding the ground state see, for example,  Ref. \cite{JENKINS}.

The extension of the $1/N_c$ expansion method to excited states 
requires the symmetry group  SU($2N_f$) $\times$ O(3) \cite{Goi97}, in order to introduce orbital excitations.
The practice shows that the experimentally observed resonances can approximately be classified  
as  SU($2N_f$) $\times$ O(3) multiplets, grouped into 
excitation bands, $N $ = 1, 2, 3, ..., 
each band containing a number of SU(6) $\times$ O(3) multiplets,
as in quark models. In addition, lattice QCD studies 
have shown that the number of each spin and flavor states in the lowest energy bands is in agreement
with the expectations based on a weakly broken SU(6) $\times$ O(3) symmetry \cite{Edwards:2012fx}, used in quark models and in the 
treatment of excited states in large $N_c$ QCD. Presently the lattice QCD report errors bars on the baryon 
masses larger than the next order corrections in the mass formula of the 
$1/N_c$  expansion  \cite{Fernando:2014dna}.

Some symmetric multiplets of SU(6) $\times$ O(3), in particular $[{\bf 56}, 2^+]$ and $[{\bf 56}, 4^+]$,
containing two and four units of orbital excitations,
were analyzed by analogy to the ground state in Refs. \cite{GSS03} and \cite{Matagne:2004pm} respectively.
In this case the splitting starts at order $1/N_c$ as well.

For mixed symmetric states the situation is more intricate. Two approaches 
have been proposed so far. The first one is based on the Hartree approximation and
describes the $N_c$ quark system as a ground state symmetric core of $N_c - 1 $ 
quarks and an excited quark \cite{CCGL}. This implies the split of 
SU($2N_f$) generators into two parts, one acting on the  core and the other on the excited quark.
Naturally, the number of generators entering the mass formula becomes larger, hence the applicability
of the method beyond the $N$ = 1 band becomes more problematic  \cite{Matagne:2006zf}.

The second procedure, where the Pauli principle is
implemented to all $N_c$ identical quarks has been proposed in Refs. \cite{Matagne:2006dj,Matagne:2008kb}.
There is no physical reason to separate the excited quark from the rest of the system.
The method can straightforwardly be applied to all excitation bands $N$. It requires 
the knowledge of the matrix elements of all the  SU($2N_f$) generators acting on mixed 
symmetric states described by the partition $(N_c - 1,1) $.
In both cases the mass splitting starts at order $N^0_c$. The latest achievements for the ground state and the 
current status of large $N_c$ excited baryons can be found in Ref. \cite{Matagne:2014lla}.

The present work considers as an example the mixed symmetric $[{\bf 70, \ell^+}]$ multiplet in the spirit of 
the procedure of Refs. \cite{Matagne:2006dj,Matagne:2008kb}. This multiplet has already been analyzed in
Ref.  \cite{Matagne:2016gdc} by using the 2014 version of the Review of Particle Properties (PDG2014) \cite{PDG}.
We use the same formalism as in Ref.  \cite{Matagne:2016gdc} but propose a new assignment to the 
$\Lambda(2110)5/2^+$*** resonance. Here we suggest that it belongs to the  quartet 
$^4\Lambda[{\bf 70},2^+]\frac{5}{2}^+$ instead of the  
$^2\Lambda[{\bf 70},2^+]\frac{5}{2}^+$ doublet. In addition, for its experimental mass we use the average value of the
2016 Review of Particle Properties (PDG2016) \cite{Olive:2016xmw}
instead of the mass found by Zhang et al. \cite{Zhang:2013sva}. This cures the previous anomaly that in some sectors 
the hyperon $\Lambda$ appears with a smaller mass than the nucleon partner \cite{Matagne:2016gdc}.
As a benefit, predictions for a few unknown hyperons are made. 
   
In Sec. \ref{massformula} we recall the mass formula of the $1/N_c$ expansion
and in Sec. \ref{review} we shortly review the applications of the method to $N$ = 1, 2, 3 and 4
excitation bands. The matrix elements of the SU(3) flavor symmetry breaking 
operators $B_i$ for the mixed symmetric $[{\bf 70, \ell^+}]$ multiplet of the
$N$ = 2 band are presented in Sec. \ref{matrixelements}. The spectrum of $[{\bf 70, \ell^+}]$
is reanalyzed in Sec.  \ref{fit}.  
The last section is devoted to  conclusions. 


\section{Mass Operator}\label{massformula}

The general form of the mass operator,  where the SU(3) symmetry is broken, has the following form \cite{JL95} 
\begin{equation}
\label{massoperator}
M = \sum_{i}c_i O_i + \sum_{i}d_i B_i .
\end{equation} 
The rotational invariant operators $O_i$ are defined as the scalar products
\begin{equation}\label{OLFS}
O_i = \frac{1}{N^{n-1}_c} O^{(k)}_{\ell} \cdot O^{(k)}_{SF},
\end{equation}
where  $O^{(k)}_{\ell}$ is a $k$-rank tensor in SO(3) and  $O^{(k)}_{SF}$
a $k$-rank tensor in SU(2)-spin, but invariant in SU($N_f$).
For the ground state one has $k = 0$. The excited
states also require  $k = 1$  and $k = 2$ terms.
The $k = 1$ tensor components are the generators $L^i$ of SO(3). In a spherical basis
the components of the $k = 2$ tensor operator of SO(3) ($i,j$ = - 1, 0, 1) read (see Appendix)
\begin{equation}\label{TENSOR} 
L^{(2)ij} = \frac{1}{2}\left\{L^i,L^j\right\}-\frac{1}{3}(-)^i
\delta_{i,-j}\vec{L}\cdot\vec{L}.
\end{equation}
The operators  $O^{(k)}_{SF}$ are constructed from the SU($N_f$) generators,
$S^i$,  $T^a$ and $G^{ia}$ obeying the 
su(2$N_f$) algebra 
\begin{widetext}
\begin{eqnarray}\label{ALGEBRA}
&[S^i,S^j]  =  i \varepsilon^{ijk} S^k,
~~~~~[T^a,T^b]  =  i f^{abc} T^c,   ~~~~~~ [S^i,T^a] = 0,   \nonumber \\
&[S^i,G^{ja}]  =  i \varepsilon^{ijk} G^{ka},
~~~~~[T^a,G^{jb}]  =  i f^{abc} G^{jc}, \nonumber \\
&[G^{ia},G^{jb}] = \fr{i}{4} \delta^{ij} f^{abc} T^c
+\frac{i}{2} \varepsilon^{ijk}\left(\frac{1}{N_f}\delta^{ab} S^k 
+d^{abc} G^{kc}\right),
\end{eqnarray}
\end{widetext}

In the symmetric core + excited quark procedure \cite{CCGL}
each SU(2$N_f$) generator is split into two parts   
\begin{equation}\label{coreplusquark}
S^i = S^i_c + s^i, T^a = T^a_c + t^a, G^{ia} =   G^{ia}_c + g^{ia},
\end{equation}
where the operators carrying a lower index $c$ act on a symmetric ground state core
and $s^i$, $t^a$ and $g^{ia}$ act on the excited quark.
The procedure has the algebraical advantage that it reduces the problem of the knowledge of 
the matrix elements of the SU($2N_f$) generators acting on a system
described by a  mixed symmetric representation of  SU($2N_f$)
to the knowledge of the matrix elements of $S^i_c$,  $T^a_c$ and  $G^{ia}_c$, acting on symmetric states
of partition $[N_c - 1]$,
which are simpler to find than the matrix elements of the SU($2N_f$) generators for
$[N_c - 1,1]$ mixed symmetric states. 
Then the operator reduction rules for the ground state \cite{DJM95}
may be used for the core operators. However, the number of  terms  to 
be included in operators describing observables remains usually very large
as compared to  experimental data, so that the method cannot easily be applied 
to  mixed symmetric highly excited baryons. It should be remembered
that  the spin-orbit operator $O_2$ of symmetric multiplets
is defined in terms of angular momentum $L^i$ components acting on the whole
system as in Ref. \cite{GSS03} and is order  $\mathcal{O}(1/N^c)$ 
\begin{equation}\label{LS}
 O_2 = \frac{1}{N_c} L \cdot S,   
\end{equation}
while for mixed symmetric multiplets it is defined as a single-particle operator  \cite{CCGL}
\begin{equation}\label{newspinorbit}
O_2 = \ell \cdot s = \sum^{N_c}_{i=1} \ell(i) \cdot s(i),
\end{equation}
the matrix elements of which are  order  $\mathcal{O}(N^0_c)$. The reason to mention $O_2$
is that although its contribution to the mass is generally small, like in quark models, here
it plays an important role in proving the compatibility between the meson-nucleon scattering
picture and the quark model-type picture, legitimating in this way the extension of the 
$1/N^c$ expansion to excited states of mixed symmetry \cite{COLEB1}.

An extra complication for $N_f$ = 3 ($u$, $d$, $s$ quarks) is that the effects of
the  SU(3) flavor symmetry breaking are comparable to $1/N_c$ corrections. The second 
term in the mass formula (\ref{massoperator}) is designed to introduce the symmetry breaking.
The operators $B_i$ break the SU(3) flavor symmetry and are defined to have zero expectation
values for nonstrange baryons. The SU(3) flavor symmetry breaking 
is implemented at order  $\mathcal{O}(\epsilon)$ where $\epsilon$ $\sim$ 0.3 is a measure of the
SU(3) flavor symmetry breaking by the strange quark mass \cite{JL95}. Thus  $\epsilon$
and $1/N_c$ at $N_c$ = 3 are of similar size and both corrections have to be included.
Corrections of order  $\epsilon/N_c$ are neglected.

In the context of our approach, where the baryon is treated as a system of $N_c$ 
quarks irrespective of its spin-flavor symmetry, the SU(3) breaking operators are defined as 
\begin{equation}\label{operatorB1}
B_1 = n_s,
\end{equation}
where $n_s$ is the number of strange quarks and  
\begin{equation}\label{operatorB2}
B_2 = \frac{1}{N_c}(L^i G^{i8} - \frac{1}{2 \sqrt{3}} L \cdot S),
\end{equation}
\begin{equation}\label{operatorB3}
B_3 =\frac{1}{N_c}(S^iG^{i8} - \frac{1}{2 \sqrt{3}} S\cdot S),
\end{equation}
where the angular momentum operator $L^i$, the spin operator $S^i$ and the component 8 of the 
spin-flavor operator $G^{i8}$ act on the entire system of  $N_c$ quarks.

Then, in Eq. (\ref{massoperator}) the coefficients $c_i$ encode the quark dynamics and $d_i$ 
measure the SU(3) breaking. They are determined from a numerical fit to data. An example,
containing the commonly used $O_i$ and $B_i$ operators together with the 
coefficients  $c_i$ and $d_i$ can be found in Table \ref{operators} below.

\section{Status of excited hyperons in the $1/N_c$ expansion}\label{review}

Here we briefly recall some important achievements in the study of baryons spectra for the 
$N$ = 1, 2, 3 and 4 bands. 

\subsection{$N$ = 1 band}

The $N$ = 1 band has been the most studied so far. It is the best known experimentally
and it contains only one SU(6) $\times$ O(3) multiplet, the $[{\bf 70, 1^-}]$. The first application 
of the $1/N_c$ expansion was a phenomenological analysis of strong decays of resonances
with one unit of orbital excitation \cite{Carone:1994tu}. There were no operators to
distinguish the strange quark from $u$ and $d$, but the decay of some hyperons was considered
via an explicit SU(3)-flavor  breaking.

In the symmetric core + excited quark procedure, the $N_f$ = 3 case has been 
thoroughly studied  by Goity et al. \cite{Goity:2002pu} where 11 SU(3)
exact flavor symmetry operators and 4 first order SU(3)-flavor symmetry  breaking operators
operators were included. Two of them, proportional to the generators $t^8$ and $T^8_c$,
thus giving a measure of the strangeness, bring significant contributions,
the other two bring  small contributions. 
The fit was made to
19 empirical quantities (17 masses and 2 mixing angles) associated to three- and
four-star resonances. Predictions were made for unknown hyperons having strangeness $\mathcal{S}$ =
- 1, - 2 and - 3. The masses of $\Lambda$(1405) and $\Lambda$(1520) were well reproduced,
but this was due to the simplicity of the wave function in the symmetric core + excited quark procedure
where the  part corresponding 
to $S_c$ = 1 is missing \cite{Matagne:2014lla}.
In addition one should note the absence of the pure flavor operator $t \cdot T_c$
coupling the core flavor operator $T_c$ to the excited quark flavor operator $t$.

A much smaller number of operators was needed for the $[{\bf 70, 1^-}]$ multiplet in the
approach of Refs.  \cite{Matagne:2006dj,Matagne:2008kb}. There were seven exact SU(3)-flavor symmetry,
one   SU(3)-flavor symmetry  breaking representing the total strangeness  and one isospin breaking operator. 
This approach, based on an exact wave function, accommodates a slightly heavier $\Lambda(1405)$ at 
1421$\pm$14 MeV. However, both procedures predict too large a  mass (of 1790 MeV in Ref. \cite{Matagne:2011fr})
for the three-star puzzling $\Xi(1690)$ resonance, a situation similar to quark models \cite{Pervin:2007wa}. 
The Skyrme model gives a lower mass and 
possibly a more natural interpretation of $\Xi(1690)$ \cite{Oh:2007cr}.

We note that in both approaches the $\Lambda$ - $N$ splitting is similar, around 150 MeV for octets. 
In decuplets the $\Sigma$ - $\Delta$ splitting is about 130 MeV in Ref. \cite{Goity:2002pu}
and about 170 MeV in Ref. \cite{Matagne:2011fr} where a different choice of $B_i$ operators
has been made, as implied by arguments given in the Introduction.

\subsection{$N$ = 2 band}

The $N$ = 2 band has the following multiplets
$[{\bf 56'},0^+]$,    $[{\bf 56},2^+]$, $[{\bf 70},0^+]$,  $[{\bf 70},2^+]$ and 
$[{\bf 20},1^+]$. The observed resonances are usually assigned to the symmetric [{\bf 56}] or the mixed symmetric $[{\bf 70}]$
SU(6) multiplets. The antisymmetric SU(6)$\times$ O(3)  multiplet $[{\bf 20},1^+]$ has been ignored so far,
on the basis that it does not have a real counterpart.

The multiplet  $[{\bf 56'},0^+]$ describes states  with a radial excitation,
in particular the Roper resonance. It was the first to be studied in the large $N_c$  limit \cite{Carlson:2000zr},
by using a simplified mass formula of the  G\"ursey-Radicati type. The analysis was free of any assumption regarding the
wave function except its symmetry in SU(6). Strong decay widths were calculated as well.

The analysis of the  $[{\bf 56},2^+]$ baryon masses has first been  performed in Ref. \cite{GSS03}. It has been 
reconsidered in Ref. \cite{Matagne:2004pm} with nearly identical results and  the analysis has been extended to the higher multiplet   
$[{\bf 56},4^+]$ of the $N$ = 4 band in the same paper.

The $[{\bf 70},0^+]$ and $[{\bf 70},2^+]$  baryon masses were first analyzed in Ref. \cite{Matagne:2005gd} for $N_f$ = 2 and
extended in Ref. \cite{Matagne:2006zf} to $N_f$ = 3, both studies being performed within the symmetric core + excited
quark procedure \cite{CCGL}. 
The $[{\bf 70},\ell^+]$ ($\ell$ = 0, 2) multiplets were revisited \cite{Matagne:2013cca}
within the approach of Ref. \cite{Matagne:2006dj}
where the Pauli principle was fully taken into account.

In Refs. \cite{Matagne:2005gd} and \cite{Matagne:2013cca} Regge-type trajectories have been drawn for the most dominant
coefficient in the mass formula, $c_1$ and $c_1^2$ respectively, and somewhat conflicting results have been obtained.
The trajectories were derived as a function of the band number $N = 0,1,2,3$ and 4.
While in Ref. \cite{Matagne:2005gd} a single trajectory has been obtained (note that large $N_c$ results for the $N$ = 3 band 
were not  available yet), in Ref. \cite{Matagne:2013cca} two distinct, nearly parallel, Regge trajectories have been obtained,
the lower one for symmetric $[{\bf 56}]$-plets and the higher one for mixed symmetric $[{\bf 70}]$-plets.

In Ref.  \cite{Matagne:2016gdc}
a combined analysis of the $[{\bf 56}, 2^+]$ and $[{\bf 70}, \ell^+]$
multiplets of the $N$ = 2 band has been made. An important aspect was that the same set of linearly independent
operators in the mass formula  has been used which was not the case before. Distinct Regge trajectories resulted again.
The data were from PDG2014 
which sometimes gives more precise values for the resonance masses with smaller error bars than before.

\subsection{$N$ = 3 and 4 bands}\label{3and4}

The $N$ = 3 band contains eight SU(6) $\times$ O(3) multiplets \cite{Stancu:1991cz}. Those 
belonging  to the mixed symmetric $[{\bf 70},\ell^-]$ multiplets ($\ell$ = 1,2,3)
were studied in Ref. \cite{Matagne:2012tm}. They were all nonstrange baryons. It is premature
to perform an extended $1/N_c$ analysis to the $N$ = 3 band, due to lack of experimental data.

The $N$ = 4 band has 17
SU(6) $\times$ O(3) multiplets \cite{Stassart:1997vk} from which only the the lowest, the $[{\bf 56},4^+]$
multiplet, has been analyzed in the $1/N_c$ expansion method \cite{Matagne:2004pm}. 
Being described by a symmetric representation of SU(6) it is
technically simple, as mentioned in the Introduction.
Despite the lack
of data for highly excited hyperons,  tentative predictions have been made in 
Ref.  \cite{Matagne:2004pm} by including only $B_1$
and a single experimentally known hyperon, the $\Lambda(2350)9/2^{+***}$.


\section{Matrix elements of $B_i$ operators for $[\bf 70,\ell^+]$}\label{matrixelements}

Here we are concerned with the $[\bf 70,\ell^+]$ multiplet.
The  matrix elements of $O_i$ for $[\bf 70,\ell^+]$,
as a function of $N_c$, were derived in Ref. \cite{Matagne:2013cca}.
Note that in the case of mixed symmetric states  the matrix elements of  $O_6$ are  $\mathcal{O}(N^0_c)$,
in contrast to the symmetric case where they are $\mathcal{O}(N^{-1}_c)$, and  nonvanishing only
for octets, while for the symmetric case they are nonvanishing for decuplets. 
Thus, at large $N_c$ the splitting starts at order $\mathcal{O}(N^{0}_c)$ for mixed symmetric states 
due both to $O_2$ and $O_6$.

The SU(3) flavor breaking operators $B_i$ were chosen to have identical definitions 
for mixed symmetric  multiplets \cite{Matagne:2016gdc}
to those for symmetric multiplets \cite{GSS03}.
 The expectation value of $B_1$ is 
\begin{equation}\label{B1}
B_1 = n_s
\end{equation}
where $n_s$ is the number of strange quarks in a baryon.
The diagonal matrix elements of $B_2$ and $B_3$  for $[\bf 70,\ell^+]$ at arbitrary $N_c$ were first 
calculated in Ref.  \cite{Matagne:2016gdc} where they were exhibited in Table IV.  
For practical purposes we do not reproduce that table. At $N_c$ = 3 we have summarized
those results  by two simple analytic formulas.  
The diagonal matrix elements of $B_2$ take the following form
\begin{equation}\label{B2}
  B_2 = - n_s 
\frac{\langle L \cdot S \rangle}{6 \sqrt{3}},
\end{equation}
where 
${\langle L \cdot S \rangle}$ is the expectation value of the spin-orbit operator acting on the whole system.
Thus the  contribution of  $B_2$ is positive or negative depending on the sign of ${\langle L \cdot S \rangle}$.
The diagonal matrix elements of $B_3$ take the simple analytic form 
\begin{equation}\label{B3}
  B_3  = - n_s
\frac{S(S + 1)}{6 \sqrt{3}},
\end{equation}
where  $S$ is the total spin. The contribution of $B_3$ is always negative, otherwise vanishing 
for nonstrange baryons.
These formulas can be applied to  $^28_J$, $^48_J$, 
$^2{10}_J$ and $^2{1}_{1/2}$ baryons of the  $[{\bf 70},\ell^+]$ multiplet. 

Interestingly, for the decuplet members of the symmetric $[{\bf 56}, 2^+]$ multiplet
the expressions of $B_2$ and $B_3$ at $N_c$ = 3
given by Eqs. (12) and (13) of Ref.  \cite{Matagne:2016gdc}
are the same as those of Eqs. (\ref{B2}) and  (\ref{B3}) shown above.

\section{Spectrum of $[{\bf 70},\ell^+]$}\label{fit}

Presently we use the PDG2016 \cite{Olive:2016xmw} to reanalyze
the mixed symmetric multiplet $[{\bf 70},\ell^+]$ with $\ell$ = 0 or 2.
The values of the fitted coefficients $c_i$ and
$d_i$ are exhibited in  Table \ref{operators} together with the value of $\chi_{\mathrm{dof}}^2$ = 1.80.
The results  can only roughly be compared to those presented in Table I, Fit 2
of Ref. \cite{Matagne:2013cca}, because  $B_2$ and $B_3$ were missing there. Note that the
factor 15 of $O_6$ has been removed here, which explains the larger value of $c_6$ now.
In fact the product $c_6 O_6$ matters in the mass. The value of $c_2$ is similar to that of
Ref. \cite{Matagne:2013cca}. 
The $1/N_c$ corrections
are dominated by $O_3$ in octets and by $O_4$ in decuplets. The SU(3) flavor breaking 
is dominated by $B_1$ for all hyperons.

\begin{table}[htb]
\caption{List of dominant operators and their coefficients (MeV) $c_i$ and $d_i$
from the mass formula (\ref{massoperator}) obtained 
in a numerical fit for the $[{\bf 70},\ell^+]$ multiplet.
The spin-orbit operator $O_2$ is defined by 
Eq.(\ref{newspinorbit}) for $[{\bf 70},\ell^+]$. }
\label{operators}
\renewcommand{\arraystretch}{1.2} 
\begin{tabular}{lrrrr}
\hline
\hline
Operator & &
Coefficient(MeV) \hspace{0.01cm}  &\\
\hline
\hline
$O_1 = N_c \ \1 $     &  & 
   630 $\pm$ 11 & \\
$O_2$ = $\ell \cdot s$ &  &                 
 62 $\pm$ 26 &    \\
$O_3 = \frac{1}{N_c}S^iS^i$  &  &   
95 $\pm$ 31 &     \\
$O_4 = \frac{1}{N_c}\left[T^aT^a-\frac{1}{12}N_c(N_c+6)\right]$ &  &   108 $\pm$ 43 &   \\[0.8ex]
$O_6 =  \frac{1}{N_c} L^{(2)ij} G^{ia} G^{ja}$ & &
    137 $\pm$ 57 &    \\[0.5ex]
\hline
$B_1 = n_s$                                     & 
&     40 $\pm$ 33 &      \\ 
$B_2 = \frac{1}{N_c}(L^iG^{i8}  - \frac{1}{2\sqrt{3}} L^i S^i)$    &   
& - 37 $\pm$ 122 & \\
$B_3 = \frac{1}{N_c}(S^iG^{i8}  - \frac{1}{2\sqrt{3}} S^i S^i)$    & 
&    60 $\pm$ 162 &  \\[0.8ex]
\hline                  
$\chi_{\mathrm{dof}}^2$                                            &     
&    1.80 &    \\

\hline \hline
\end{tabular}
\end{table}

The PDG2016 as well as PDG2014 
incorporate the new 
multichannel partial wave analysis of the Bonn-Gatchina group  \cite{Anisovich:2011fc}. 
Accordingly 
the resonance $P_{13}(1900)$ has been upgraded from two to three stars with a Breit-Wigner mass of
1905 $\pm$  30 MeV.
The resonance  $N(2000)5/2^+$ 
has been split into two two-star resonances, namely $N(1860)5/2^+$ and $N(2000)5/2^+$,
with masses  indicated in Table \ref{MASSES}.
There is a new one-star resonance $N(2040)3/2^+$ observed in the decay $J/\psi \rightarrow p \bar p \pi^0$
\cite{Ablikim:2009iw}.
There is also a new two-star resonance $N(1880)1/2^+$  observed by the Bonn-Gatchina group with a mass of
1870 $\pm$ 35 MeV \cite{Anisovich:2011fc}.

In a previous work  \cite{Matagne:2013cca}  only 11 resonances have been included in the numerical fit. 
Here, as well as in Ref.   \cite{Matagne:2016gdc}, 
16 resonances have been included, with a status of three, two or one star. 
These extra resonances are the  hyperons
$\Xi(2120)?^{?*}$, $\Sigma(2070)5/2^{+*}$, $\Sigma(1940)?^{?*}$,  $\Xi(1950)?^{?***}$ and
$\Sigma(2080)3/2^{+**}$. 
For the three-star resonances we use the Breit-Wigner mass of PDG2016
except for $\Xi(1950)?^{?***}$ where we take the value found in Ref. \cite{Biagi:1986vs} which reduces the 
$\chi_{\mathrm{dof}}^2$   value from 1.96 to 1.80. For the spectrum, such a choice would not make much difference.

For the resonances omitted from the summary table of PDG2016
the masses and the error bars considered
in the fit correspond to averages over those data taken into account in the particle listings,
except for a few which favor specific experimental values cited in the headings of Table \ref{MASSES}.

The $N(1710){1/2^{+***}}$ and
$\Sigma(1770){1/2^{+*}}$ resonances have been ignored in this fit.
The theoretical argument is that their masses are too low, leading to unnatural sizes for the 
coefficients  $c_i$ or $d_i$ \cite{Pirjol:2003ye}.
Experimentally  the controversial $N(1710)1/2^{+***}$ resonance has not been seen in
the latest GWU analysis  of Arndt et al. \cite{Arndt:2006bf}. 
We have also ignored $\Delta(1750)1/2^{+*}$,
inasmuch as,
neither Arndt et al. \cite{Arndt:2006bf} nor Anisovich et al.  \cite{Anisovich:2011fc} find evidence
for it.

{\squeezetable
\begin{table}
\caption{Partial contribution and the total mass (MeV) predicted by the $1/N_c$ expansion
with matrix elements of $O_i$ from Ref. \cite{Matagne:2016gdc} and of $B_i$ given in the text.
The column  Ref.[24] gives the total mass of Ref.  \cite{Matagne:2016gdc}.
The last two columns give the empirically known masses and status from the 2016 Review of Particles Properties 
\cite{Olive:2016xmw} unless specified by (A) from \cite{Anisovich:2011fc}, 
(L) from \cite{Litchfield:1971ri},
(G1) from  \cite{Gopal:1980ur}, (B) from \cite{Biagi:1986vs},
(AB) from \cite{Ablikim:2009iw},
(G2) from  \cite{Gopal:1976gs}
.}\label{MASSES}
\renewcommand{\arraystretch}{1.5}
\begin{tabular}{crrrrrrccccccl}\hline \hline
                    &      \multicolumn{8}{c}{Part. contrib. (MeV)}  & \hspace{0.0cm} Total(MeV) & Ref.[24] & \hfill Expt.(MeV)\hspace{0.0cm}
 &\hspace{0.cm}  Name, status \hspace{.0cm} \\

\cline{2-9}
                    & \hspace{.0cm} $c_1O_1$  & \hspace{.0cm}  $c_2O_2$ & \hspace{.0cm}$c_3O_3$ &\hspace{.0cm}  $c_4O_4$ 
&\hspace{.0cm}  $c_6O_6$ & $d_1B_1$ & $d_2B_2$ & $d_3B_3$ &  \\
\hline
$^4N[{\bf 70},2^+]\frac{7}{2}^+$        & 1889 & 62 & 118 & 27 & - 23 & 0  &  0  &    0  & $2073\pm38$ & $2080\pm39$ & $2060\pm65$(A) & $N(1990)7/2^+$**  \\
$^4\Lambda[{\bf 70},2^+]\frac{7}{2}^+$  &       &    &     &    &     & 40 & 11  & - 22  & $2102\pm19$ & $2105\pm19$ & $2100\pm30$(L) 
& $\Lambda(2020)7/2^+$* \\
$^4\Xi[{\bf 70},2^+]\frac{7}{2}^+$      &       &    &     &    &      & 79 & 22  & - 43 & $2131\pm 8$ & $2130\pm 8$ &
$2130\pm8$  &  $\Xi(2120)?^?$* \vspace{0.2cm}\\
\hline
$^4N[{\bf 70},2^+]\frac{5}{2}^+$        & 1889 & - 10  & 118 & 27 & 57 & 0  & 0   & 0    & $2081\pm33$ & $2042\pm41$ &$2000\pm50$ & $N(2000)5/2^+$**\\
$^4\Lambda[{\bf 70},2^+]\frac{5}{2}^+$  &      &       &     &    &    & 40 & - 2 & - 22 & $2097\pm18$ & $2009\pm40$ & $2110\pm20$ & $\Lambda(2110)5/2^+$***\\
$^4\Xi[{\bf 70},2^+]\frac{5}{2}$        &      &       &     &    &    & 79 & - 4 & - 43 & $2113\pm41$   &              & & \vspace{0.2cm}\\
\hline
$^4N[{\bf 70},2^+]\frac{3}{2}^+$        & 1889 & - 62  &  118 & 27  & 0  &  0  &    0  & 0     & $1972\pm29$  & $1955\pm32$    &    &  \\
$^4\Lambda[{\bf 70},2^+]\frac{3}{2}^+$  &      &       &      &     & 0  & 40  & - 11  & - 22  & $1979\pm42$       & \\
$^4\Xi[{\bf 70},2^+]\frac{3}{2}$        &      &       &      &     &    & 79  & - 22  & - 43  & $1986\pm99$       &  \\
\hline
$^4N[{\bf 70},2^+]\frac{1}{2}^+$        & 1889 &- 93 & 118 & 27  & - 80 & 0  &  0   & 0    & $1861\pm33 $   & $1878\pm34$ & $1870\pm35$(A) & $N(1880)1/2^+$**\\
$^4\Lambda[{\bf 70},2^+]\frac{1}{2}^+$  &      &      &     &    &    & 40 & - 16 & - 22 & $1863\pm79 $ &                & \\ 
$^4\Xi[{\bf 70},2^+]\frac{1}{2}^+$      &      &      &     &    &    & 79 & - 32 & - 43 & $1865\pm153 $ &           & \vspace{0.2cm}\\              
\hline
$^2N[{\bf 70},2^+]\frac{5}{2}^+$        & 1889 & 21  & 23  & 27  & 0   & 0   &  0 &  0   & $1960\pm29$  & $1959\pm29$& $1860\pm{^{120}_{60}}$(A) & $N(1860)5/2^+$** \\
$^2\Sigma[{\bf 70},2^+]\frac{5}{2}^+$   &      &     &     &      & 0  & 40  &  4 & - 4 & $2000\pm18$ & $2031\pm11$&  $2051\pm25$(G1) 
& $\Sigma(2070)5/2^+$* \vspace{0.2cm}\\
$^2\Xi[{\bf 70},2^+]\frac{5}{2}$        &      &     &     &     &     & 79  &  7 & - 8 & $2038\pm45$ &   \vspace{0.2cm}\\
\hline
$^2N[{\bf 70},2^+]\frac{3}{2}^+$        & 1889 &- 31 & 23  & 27  & 0   &  0  & 0  & 0   & $1908\pm21$ & $1902\pm22$ & $1900\pm30$(A) & $N(1900)3/2^+$***  \\
$^2\Sigma[{\bf 70},2^+]\frac{3}{2}^+$   &      &     &     &     & 0   &  40 & - 6 & - 4 & $1938\pm16$ & $1933\pm11$ & $1941\pm18$ & $\Sigma(1940)?^?$*  
\vspace{0.2cm} \\
$^2\Xi[{\bf 70},2^+]\frac{3}{2}^+$      &      &     &     &     & 0   &  79 & - 11 & - 8 & $1968\pm7$ & $1964\pm70$ & $1967\pm7$(B) & $\Xi(1950)?^?$*** \vspace{0.2cm}\\
\hline
$^4N[{\bf 70},0^+]\frac{3}{2}^+$         & 1889 &  0  & 118 & 27  &  0  &  0  & 0 & 0      & $2034\pm18$  & $2024\pm20$
& $2040\pm28$(AB)  & $N(2040)3/2^+$*\\
$^4\Sigma[{\bf 70},0^+]\frac{3}{2}^+$    &      &     &     &     &     &  40 & 0 & - 22  & $2052\pm22$  & $2000\pm23$ & $2100\pm69$   
& $\Sigma(2080)3/2^+$** \\
$^4\Xi[{\bf 70},0^+]\frac{3}{2}^+$       &      &     &     &     &     &  79 & 0 & - 43  & $2070\pm46$ &  \vspace{0.2cm}\\
\hline
\hline
\end{tabular}
\end{table}}
{\squeezetable
\begin{table}
\label{multiplet}
\renewcommand{\arraystretch}{1.5}

\begin{tabular}{crrrrrrcccccl}\hline \hline
                    &      \multicolumn{8}{c}{Part. contrib. (MeV)}  & \hspace{0.0cm} Total(MeV) & Ref.[24] & \hfill  Expt.(MeV)
\hfill &\hspace{0.cm}  Name, status \hspace{.0cm} \\

\cline{2-9}
                    & \hspace{.0cm} $c_1O_1$  & \hspace{.0cm}  $c_2O_2$ & \hspace{.0cm}$c_3O_3$ &\hspace{.0cm}  $c_4O_4$ 
&\hspace{.0cm}  $c_6O_6$ & $d_1B_1$ & $d_2B_2$ & $d_3B_3$ &  \\
\hline
\hline
$^2\Delta[{\bf 70},2^+]\frac{5}{2}^+$       & 1889  & - 21  & 24 & 134  & 0  &  0  & 0 &  0  & $2026\pm48$  & $2086\pm37$ & $1962\pm139$ & 
$\Delta(2000)5/2^+$**\\
$^2\Sigma^{\ast}[{\bf 70},2^+]\frac{5}{2}^+$       &       &       &    &      & 0  &  40 & 3 & - 4  & $2065\pm52$  &  & \\
$^2\Xi^{\ast}[{\bf 70},2^+]\frac{5}{2}^+$          &       &       &    &      & 0  &  79 & 7 & - 8  & $2104\pm73$  &  & 
\vspace{0.2cm}\\
\hline
$^2\Delta[{\bf 70},0^+]\frac{1}{2}^+$       & 1889  &   0   & 24 & 134  & 0  & 0  &   0 & 0 & $2047\pm49$   &   & \\
$^2\Sigma^{\ast}[{\bf 70},0^+]\frac{1}{2}^+$&       &   0   &    &      &    & 40 & 0 & - 4 & $2083\pm46$   & $2119\pm25$  & $1902\pm96$  & $\Sigma(1880)1/2^+$**\\
$^2\Sigma^{\ast}[{\bf 70},0^+]\frac{1}{2}^+$&       &   0   &    &      &    & 79 & 0 & - 8 & $2118\pm53$   &   & 
\vspace{0.2cm}\\
\hline
$^2\Lambda'[{\bf 70},2^+]\frac{5}{2}^+$     & 1889  &  62   & 24 & - 81 & 0  & 40 & 3 & - 4 & $1933\pm47$   &  & \\
\hline
$^2\Lambda'[{\bf 70},0^+]\frac{1}{2}^+$     & 1889  &   0   & 24 & - 81 & 0  & 40 & 0 & - 4 & $1868\pm43$   & $1865\pm19$ & $1853\pm20$(G2) & $\Lambda(1810)1/2^+$*** 
\vspace{0.2cm}\\
\hline
\hline
\end{tabular}
\end{table}

The partial contributions and the calculated total masses obtained from the fit are presented in Table  \ref{MASSES}.
One can see that the fit is generally good 
except for $\Sigma(1880)1/2^{+**}$ where the calculated mass somewhat too high. The operator $B_2$ has a 
vanishing expectation value and the contribution of $B_3$, although negative, is negligible. 
The mass of the $N(1860)5/2^{+**}$ seems large too,
but it is within the large error bars of Ref. \cite{Anisovich:2011fc}.

The good fit for the $N(1880)1/2^{+**}$ resonance
was due to the negative contributions of $- 93$ MeV and $- 80$ MeV
of the spin-orbit operator $O_2$  and of $O_6$ operators respectively.  
However its strange partners are almost degenerate because the positive contribution of $B_1$  is
accidentally cancelled out by the negative contribution of $B_2 + B_3$.

The assignment of  $\Sigma(1940)?^{?*}$
and $\Xi(1950)?^{?***}$ to the $^2[{\bf 70},2^+]3/2^+$ multiplet seems reasonable. Thus these 
resonances may have $J^P$ = $3/2^+$,  hopefully  to be confirmed experimentally in future analyses. 

Some predictions have also been made for experimentally unknown strange partners in octets and decuplets.
Note that $\Lambda$ and $\Sigma$ are degenerate in our approach because the expectation
values of $B_2$ and $B_3$ are identical at $N_c$ = 3, although they are different at arbitrary $N_c$.
This is not the case for the $[{\bf 56},2^+]$ multiplet.
Also, the total contribution of $B_i$ is generally of about 30 MeV which is much less than for
the $[{\bf 56},2^+]$ multiplet.
We did not present predictions for the
$\Omega$'s in the  $[{\bf 70},\ell^+]$ multiplet because we thought them irrelevant at this stage of theory and experiment.

A useful remark is that the contributions of $B_2$ and $B_3$ mutually cancel out for hyperons belonging to 
decuplets with $\ell \ne$ 0. In that case $B_1$ is enough in the mass formula, like in Ref.  \cite{Matagne:2013cca}.
The contributions of $B_2$ and $B_3$ are generally small. This is due to the smallness of the coefficients 
$d_2$ and $d_3$ of Table \ref{operators}, having sizes of a similar order of magnitude to 
the corresponding ones from Ref.  \cite{Goity:2002pu} obtained for the $N$ = 1 band in the excited quark +
ground state core method.

Presently 
the negative contribution of $B_3$ (see Eq. (\ref{B3})) 
makes the hyperons masses
larger than those derived in  Ref.  \cite{Matagne:2016gdc} and helps in restoring the correct hierarchy as a function of strangeness. 

It is important to make a comparison between the present results and those of Ref.  \cite{Matagne:2016gdc} 
where a different assignment and mass  have been chosen for $\Lambda(2110)5/2^{+***}$. For this purpose we have
included in Table \ref{MASSES} the column called  Ref. \cite{Matagne:2016gdc} which gives the total masses 
obtained in our previous work. One can notice that presently the fit to the resonances 
$N(2000)5/2^+$** and  $\Sigma(2070)5/2^+$* slightly deteriorates, which may be a reason for the increase of 
$\chi_{\mathrm{dof}}^2$ from 1.48 to 1.80. Note that all these resonances have $J$ = $5/2^+$.

Our suggestions for assignments of resonances 
in the $[{\bf 70},\ell^+]$ multiplet  can be compared to those made in Ref. \cite{Crede:2013kia} as educated guesses. 
The assignment  of  $\Sigma(1880)1/2^{+**}$ 
as a $[{\bf 70},0^+]1/2^+$ decuplet resonance is  confirmed as well as the assignment  of $\Lambda(1810)1/2^{+***}$ 
as a flavor singlet.
We agree with Ref.  \cite{Crede:2013kia} regarding 
$\Lambda(2110)5/2^{+***}$ as a partner of $N(2000)5/2^{+**}$ in a spin quartet. 
We disagree  with Ref.  \cite{Crede:2013kia} 
that $N(1900)3/2^{+***}$ is 
a member of a spin quartet.
We propose it as a partner of $\Sigma(1940)?^{?*}$ and $\Xi(1950)?^{?***}$
in a spin doublet.

However, one has to keep in mind that at the same $J$ spin doublets
and quartets can mix, for example for $N[70,2^+]$ at $J$ = 3/2 or 5/2.
The mixing would be due to the off-diagonal matrix elements of the spin-orbit
operator $O_2$ and the tensor operator $O_6$. 

The problem of assignment is not trivial.
Within  the  $1/N_c$ expansion method Ref. \cite{GSS03} suggested that  
$\Sigma(2080)3/2^{+**}$ and $\Sigma(2070)5/2^{+*}$ could be members 
of two distinct decuplets in the $[{\bf 56},2^+]$ multiplet
while here and in Ref.   \cite{Crede:2013kia} they seem to be good candidates
for mixed symmetric states.


\section{Conclusions}

The inclusion of three SU(3) symmetry breaking operators, $B_1$, $B_2$ and $B_3$ in the mass formula
of the $[{\bf 70},\ell^+]$ multiplet helps to
brings more insight into  the SU(6) $\times$ O(3)  classification of highly excited baryons
when accompanied by realistic assignments.
Presently it seems that the evolution of the $\Lambda$ - $N$ or $\Sigma$ - $N$ splitting with excitation energy in 
baryon multiplets described by the $1/N_c$ expansion remains an open problem.

Alternative suggestions for assignments of the known baryons should be studied and more data  for excited hyperons
are highly desirable.
The continuing study of the presently available data and the production of new hyperons are needed
for understanding the structure of baryons and disentangle between various models.
At the Workshop on Physics with Neutral Kaon beam at JLAB \cite{JLAB} it was pointed out 
that a $K_L$ beam at JLAB would open new opportunities for studying excited hyperons which may
help in understanding the multiplet structure of excited baryons. Similar hopes are at J-PARK.

\appendix 

\section{The two-rank tensor of SO(3) }\label{L2}

In this Appendix we derive the expression (\ref{TENSOR}) of the two-rank tensor $L^{(2)ij}$ of SO(3)
in a spherical basis. 
Let us denote the spherical components of the SO(3) generators by $L_i$. Then the product $L_i L_j$
can be written as
\begin{equation}
L_i L_j =  \sum_{k = 0}^2  \sum_{\mu = - k}^k  C^{1 ~1 ~k}_{i ~j ~~\mu} ~T^k_{\mu},
\end{equation}
in terms of a Clebsch-Gordan coefficient and the irreducible k-rank tensor $T^k_{\mu}$.
In the anticommutator $\{L_i,L_j\}$ only the tensors $k = 0$ and 2 survive
for symmetry reasons. Then one can write
\begin{equation}\label{anti}
\frac{1}{2}\{L_i,L_j\} = \sum_{\mu} C^{1 ~1 ~k}_{i ~j ~~\mu} ~T^2_{\mu} + C^{1 ~1 ~0}_{i ~j ~~0} ~T^0_0.
\end{equation}
The second term contains the Clebsch-Gordan coefficient
\begin{equation}
C^{1 ~1 ~0}_{i ~j ~~0} = (-)^{1-i} \frac{1}{\sqrt{3}} \delta_{i,-j}.
\end{equation}
The standard definition of $T^0_0$ is (see, for example, Eq. (4.7) of Ref. \cite{BS})
\begin{equation}
T^0_0 = - \frac{1}{\sqrt{3}} \vec{L}\cdot\vec{L}.
\end{equation}
Then shifting the second term of Eq (\ref{anti}) from right to left we obtain the second rank tensor 
$L^{(2)ij}$ of SO(3) as 
\begin{equation}\label{def2}
L^{(2)ij} = \sum_{\mu} C^{1 ~1 ~2}_{i ~j ~\mu} T^2_{\mu},
\end{equation}
or alternatively
\begin{equation}\label{def1}
 L^{(2)ij} = \frac{1}{2}\{L_i,L_j\} - \frac{1}{3}(-)^i \delta_{i,-j} \vec{L}\cdot\vec{L}.
\end{equation}

Equation (\ref{def2})  can be used to calculate the matrix elements of $L^{(2)ij}$ defined as an irreducible
two-rank tensor. Using the Wigner-Eckart theorem and 
a spherical harmonic basis  one has
\begin{equation}
 \langle \ell' m' |L^{(2)ij}| \ell m \rangle =  \sum_{\mu,m} C^{1 ~1 ~2}_{i ~j ~\mu}  C^{\ell ~2 ~\ell'}_{m ~\mu ~m'}.
\langle \ell'||T^2||\ell \rangle. 
\end{equation}
The reduced matrix element $\langle \ell'||T^2||\ell \rangle$ can be easily calculated.
The result leads to
 \begin{equation}
 \langle \ell' m' |L^{(2)ij}| \ell m \rangle = \delta_{\ell' \ell} \sqrt{\frac{\ell(\ell+1)(2\ell-1)(2\ell+3)}{6}}
\sum_{\mu,m} C^{1 ~1 ~2}_{i ~j ~\mu}  C^{\ell ~2 ~\ell'}_{m ~\mu ~m'}, 
\end{equation}
which has been used in deriving the matrix elements of $O_6$ and is consistent with Eq. (A5) of Ref. \cite{CCGL}.

Equation (\ref{def1}) indicates that the definition of $L^{(2)ij}$ from Ref. \cite{Matagne:2005gd} 
contains a typographic error in the second term on the right-hand side, namely 
the phase $(-)^i$ is missing. Previous and present results are not affected by this inadvertence.


\vspace{4cm}

\centerline{\bf Acknowledgments}

This work has been supported by the Fonds de la Recherche Scientifique - FNRS under 
Grant No. 4.4501.05.


\end{document}